\documentclass[aps,prl,twocolumn,groupedaddress,amssymb,showpacs,floatfix]{revtex4}
\usepackage{graphicx}
\usepackage{psfrag}
\def\bra#1{\langle#1\vert}					
\def\ket#1{\vert#1\rangle}					
\def\bracket#1#2{\langle#1\vert#2\rangle}	
\def\ave#1{\langle#1\rangle}				

\begin{document}

\title{Stroboscopic quantum walks}

\author{O. Buerschaper}
\author{K. Burnett}
\affiliation{Clarendon Laboratory, Department of Physics, University of Oxford,
Parks Road, Oxford OX1 3PU, United Kingdom}

\date{\today}

\begin{abstract}
Discrete time (coined) quantum walks are produced by the repeated application of a constant unitary transformation to a quantum system. By recasting these walks into the setting of periodic perturbations to an otherwise freely evolving system we introduce the concept of a stroboscopic quantum walk. Through numerical simulation, we establish the link between families of stroboscopic walks and quantum resonances. These are observed in the nonlinear systems of quantum chaos theory such as the $\delta$-kicked rotator or the $\delta$-kicked accelerator.
\end{abstract}



\maketitle

Quantum walks were first mentioned in \cite{Aha93} and have been a topic of considerable interest in recent years as they exhibit properties that have led to the construction of a new class of quantum algorithms. For computationally difficult problems which can be solved classically by probabilistic algorithms \cite{Mot95} several quantum algorithms have been found which make use of quantum walks as natural counterparts of classical Markov chains. At present these new algorithms fall into two categories: algorithms achieving exponentially faster hitting times on graphs \cite{Far98, Chi02, Chi03, Kem03} and searching based on quantum walks \cite{She03, Gol03, Amb03, Amb04}. The initial discussion of quantum walks and their algorithmic applications was purely theoretical but there have also been proposals for physical realisations of such walks, using ion traps \cite{Tra02}, cold atoms in optical lattices \cite{Due02} and cavity quantum electrodynamics \cite{San03}. In real systems the quantum evolution between steps will clearly be important but this has not so far been included in the description of discrete time quantum walks. In this paper we shall outline the theoretical description of the discrete time quantum walk in one dimension and explain our generalization of the dynamics along with its implications.

The discrete time model, introduced in \cite{Amb01}, was the first scheme proposed for an abstract quantum walk. In analogy with the classical random walk a particle with internal states $\ket{0}$ and $\ket{1}$ (equivalent to heads and tails of a coin) was supposed to move on an infinite line of discrete positions which are also referred to as vertices in graph theory. In each step the walker is shifted one vertex to the right if its internal state is $\ket{0}$ and one to the left if it is $\ket{1}$. Before each translation the internal state of the particle is put into a superposition of $\ket{0}$ and $\ket{1}$ by a unitary transformation (coin toss) so that the walker is in a superposition of right and left after the swap. Repeated application of this procedure, in the absence of measurements or dissipation, produces the distinctive features, in particular enhanced diffusion, that have driven interest in these models of quantum evolution.

We now introduce a new model which we call the \emph{stroboscopic quantum walk}. The idea is that the walker, being a physical system, should also undergo time evolution and thus acquire position- and time-dependent phases, governed by the Schr{\"o}dinger equation, in between the steps of the original discrete time quantum walk. Introducing this stroboscopic scheme as a natural feature of the dynamics also casts new light on the nature of the physical description of coined quantum walks away from purely algorithmic issues. In essence we are studying the free time evolution of a quantum system in the presence of periodic perturbations. This is, of course, also a central theme in the investigation of quantum chaos \cite{Obe99, Arc01, Sch03}. Apart from being an obvious extension of coined quantum walks this free time evolution between steps in our model also drastically modifies the diffusion process. In fact it is only for special cases that the standard coined quantum walk is recovered. What is more we uncovered a whole new family of rapidly spreading walks with a variation in the rate of diffusion reminiscent of the quantum resonances of quantum chaos.

\begin{figure}
 \psfrag{X}[Bc][Bc]{$n$}
 \psfrag{Y}[Bc][Bc]{$P(n)$}
 \psfrag{Captiona}[Bc][Bc]{$\tau = 0$}
 \psfrag{Captionb}[Bc][Bc]{$\tau = T/5$}
 \psfrag{Captionc}[Bc][Bc]{$\tau = T/10$}
 \psfrag{Captiond}[Bc][Bc]{$\tau = T/(2\pi)$}
 \psfrag{A}{(a)}
 \psfrag{B}{(b)}
 \psfrag{C}{(c)}
 \psfrag{D}{(d)}
 \includegraphics{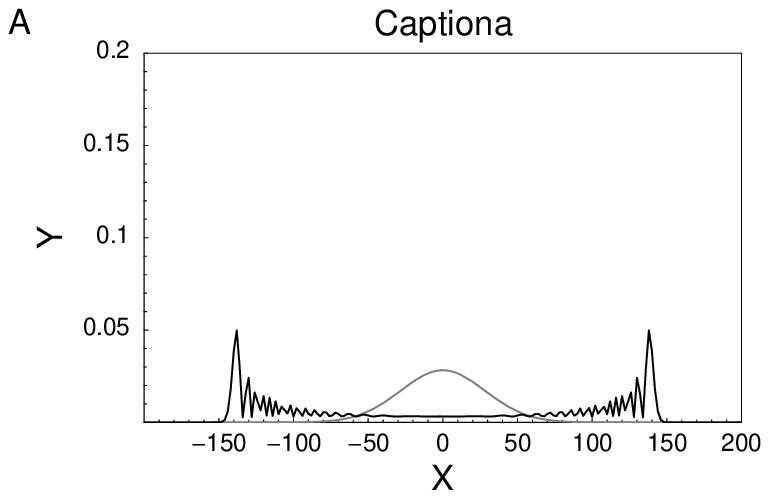}
 \includegraphics{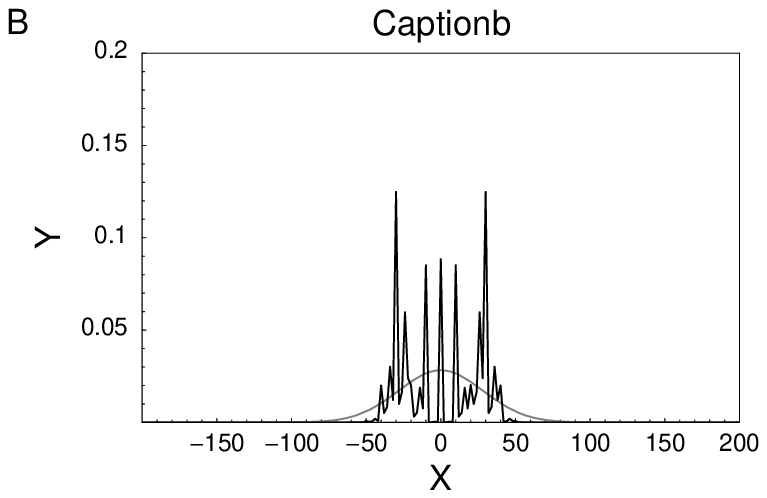}
 \includegraphics{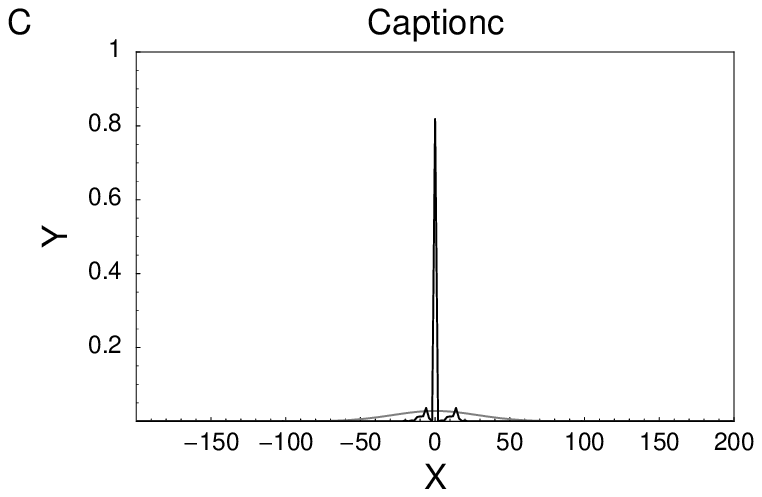}
 \includegraphics{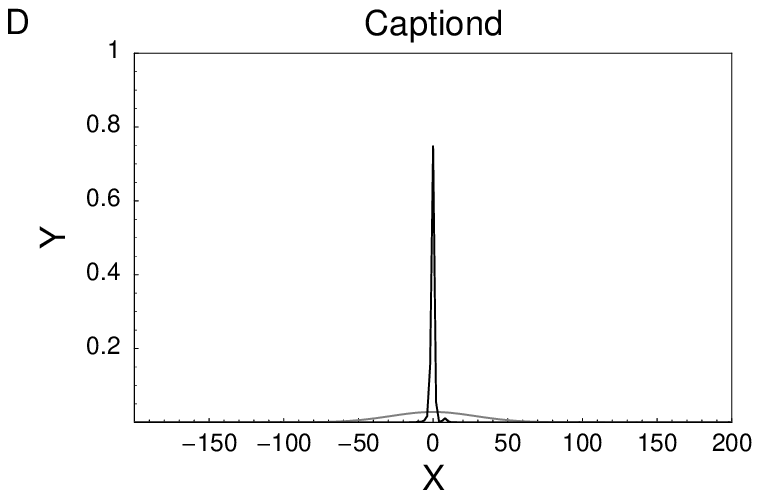}
 \caption{\label{prob} Probability distribution of the positions (black curves) after $t = 200$ steps for the ordinary Hadamard walk ($\tau=0$) as well as for stroboscopic Hadamard walks (harmonic oscillator model) with various values of $\tau$. Only even positions are drawn. For comparison the gray curves show the limiting distribution of a classical random walk.}
\end{figure}

Let us now briefly review the mathematical structure of discrete time quantum walks on the line. The total Hilbert space of the particle $\mathcal{H} = \mathcal{H}_C \otimes \mathcal{H}_P$ is spanned by $\{ \ket{0}, \ket{1} \}$ and $\{ \ket{n} \, \vert \, n \in \mathbb{Z} \}$ respectively where $\ket{n}$ denote distinct positions on the line. Hence a general state $\ket{\psi (t)}$ of the walker can be expanded in terms of these basis states, thus $\ket{\psi (t)} = \sum_{i,n} c_{in}(t) \ket{i,n}$. The coin toss $\hat{C}$ is an arbitrary unitary transformation in $\mathcal{H}_C$ which is usually taken to be the balanced Hadamard coin $\hat{H} = \frac{1}{\sqrt{2}} (\hat{\sigma}_x + \hat{\sigma}_z)$ with $\hat{\sigma}_i$ denoting the Pauli matrices in $SU(2)$. The translation operators $\hat{T}$ and $\hat{T}^{-1}$ move the particle by one vertex $\hat{T} \ket{n} = \ket{n+1}$ and $\hat{T}^{-1} \ket{n} = \ket{n-1}$. One step of the walk is thus defined as
\begin{equation}
 \hat{U} = (\ket{0} \bra{0} \otimes \hat{T} + \ket{1} \bra{1} \otimes \hat{T}^{-1}) (\hat{C} \otimes \hat{I}),
\end{equation}
repeatedly applying $\hat{U}$ to the initial state $\ket{\psi_0}$ gives the whole walk $\ket{\psi (t)} = \hat{U}^t \ket{\psi_0}$. The probability of finding the particle at vertex $n$ after $t$ steps is then given by $P(n,t) = \sum_i \vert \bracket{i,n}{\psi (t)} \vert^2$. In order to get a symmetric probability distribution on the line the initial state is chosen to be $\ket{\psi_0}= \frac{1}{\sqrt{2}} (\ket{0} + i \ket{1}) \otimes \ket{0}$, as the Hadamard coin treats $\ket{0}$ and $\ket{1}$ differently. It can be shown \cite{Amb01} that the probability distribution of a Hadamard walker has a standard deviation (i.e. expected distance from the origin) of $\sigma \propto t$ in the asymptotic limit, in other words the Hadamard walk spreads quadratically faster than its classical counterpart with $\sigma = \sqrt{t}$. The position probability distribution of the ordinary Hadamard walk is plotted in Fig.~\ref{prob} together with numerical results of some stroboscobic Hadamard walks.

We now turn to the mathematical description of stroboscopic walks. As with the discrete time model we consider a total Hilbert space of $\mathcal{H} = \mathcal{H}_C \otimes \mathcal{H}_P$ but now let the position states $\{ \ket{n} \}$ in $\mathcal{H}_P$ be eigenstates of some time-independent Hamiltonian $\hat{H}_P$ with $\hat{H}_P \ket{n} = E_n \ket{n}$. In between the steps, separated by the time interval $\tau$, the states in position space evolve according to the Schr{\"o}dinger equation. Thus we define one step of the stroboscopic walk as:
\begin{equation}
 \hat{U}(\tau) = (\ket{0} \bra{0} \otimes \hat{T} + \ket{1} \bra{1} \otimes \hat{T}^{-1}) (\hat{C} \otimes e^{-\frac{i}{\hbar} \hat{H}_P \tau}).
 \label{U}
\end{equation}
 We see that at every vertex on the line the walker effectively acquires a relative phase depending on the time interval $\tau$. If the relative phase differences between eigenstates of adjacent occupied vertices are an integer multiple of $2 \pi$ a complete rephasing occurs and no difference to the ordinary Hadamard walk will be seen. The smallest possible time interval for this to happen is called the Talbot time $T$ given by
\begin{equation}
 T = \frac{2 \pi \hbar \lambda}{E_n-E_{n+2}},
\end{equation}
where $\lambda$ is an integer depending on the energy spectrum of the particular system. Hence we can restrict our analysis to $\tau \in [0,T)$.

In order to investigate the properties of the stroboscopic walk a range of numerical simulations were performed using the simple harmonic oscillator Hamiltonian $\hat{H}_P = \hbar \omega (\hat{n} + \frac{1}{2})$ which has the Talbot time $T = \frac{\pi}{\omega}$. We start the walks at a high number state $\ket{n}$ so that the ground state $\ket{0}$ is never reached in our simulations.

\begin{figure}
 \psfrag{X}[Bc][Bc]{$t$}
 \psfrag{Y}[Bc][Bc]{$\sigma$, $\ave{n}$}
 \psfrag{Captiona}[Bc][Bc]{$\tau = 0$}
 \psfrag{Captionb}[Bc][Bc]{$\tau = T/5$}
 \psfrag{Captionc}[Bc][Bc]{$\tau = T/10$}
 \psfrag{Captiond}[Bc][Bc]{$\tau = T/(2\pi)$}
 \psfrag{A}{(a)}
 \psfrag{B}{(b)}
 \psfrag{C}{(c)}
 \psfrag{D}{(d)}
 \includegraphics{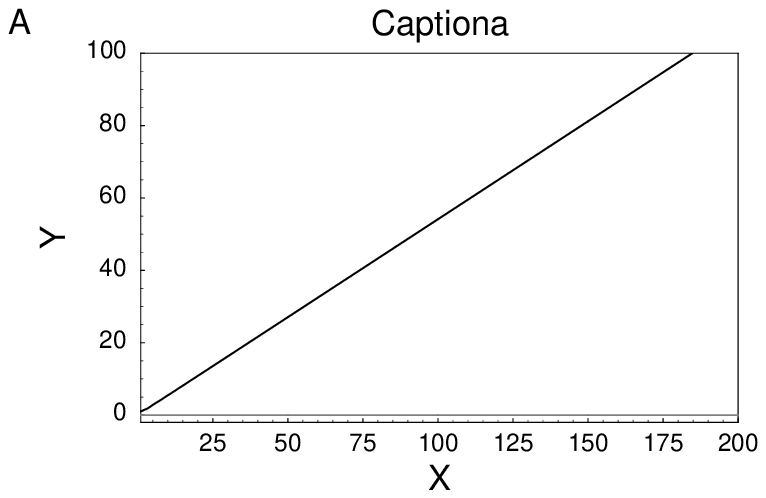}
 \includegraphics{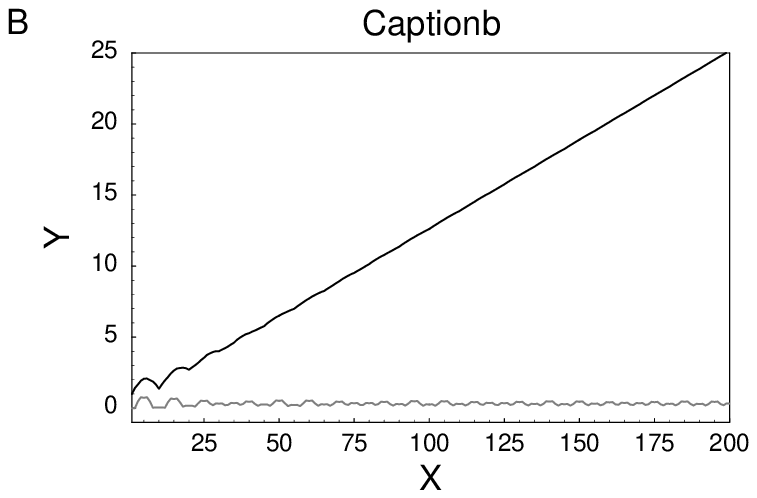}
 \includegraphics{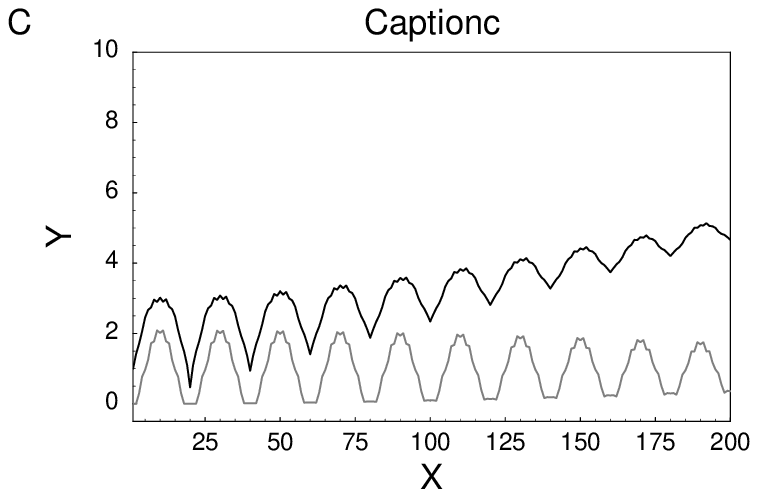}
 \includegraphics{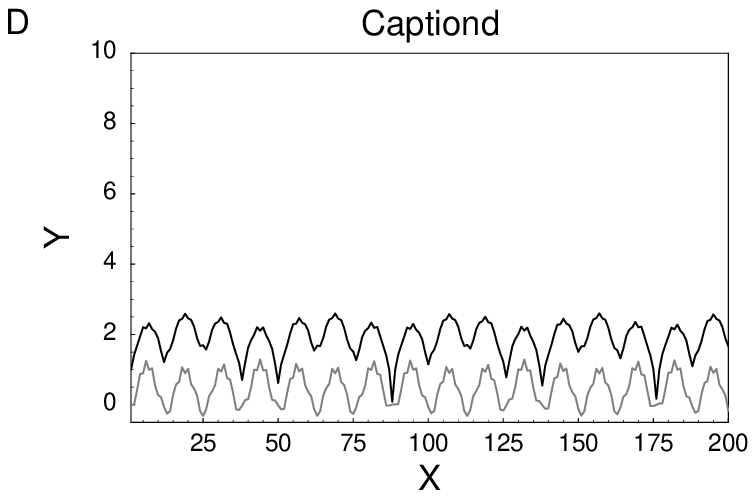}
 \caption{\label{meanvar} Standard deviation $\sigma$ (black curves) and mean $\ave{n}$ (gray curves) of the probability distribution for the ordinary Hadamard walk ($\tau=0$) as well as for stroboscopic Hadamard walks (harmonic oscillator model) with various values of $\tau$.}
\end{figure}

\begin{figure}
 \psfrag{X}[Bc][Bc]{$\tau$}
 \psfrag{Y}[Bc][Bc]{$\sigma$}
 \includegraphics{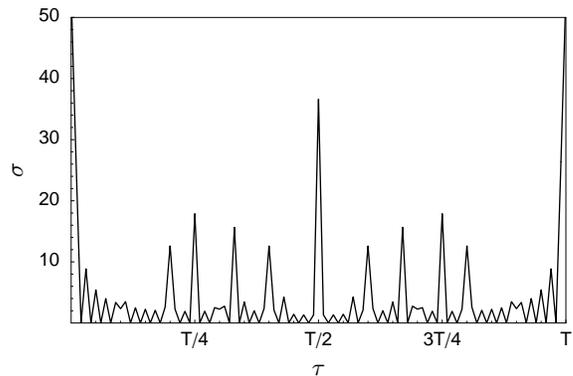}
 \caption{\label{varscan} Standard deviation $\sigma$ of stroboscopic Hadamard walks (harmonic oscillator model) after $t = 100$ steps with a time resolution of $\Delta \tau = \frac{T}{100}$.}
\end{figure}

\begin{figure}
 \psfrag{X}{$t$}
 \psfrag{Y}{$\tau$}
 \psfrag{Z}{$\frac{\sigma}{\sqrt{t}}$}
 \includegraphics{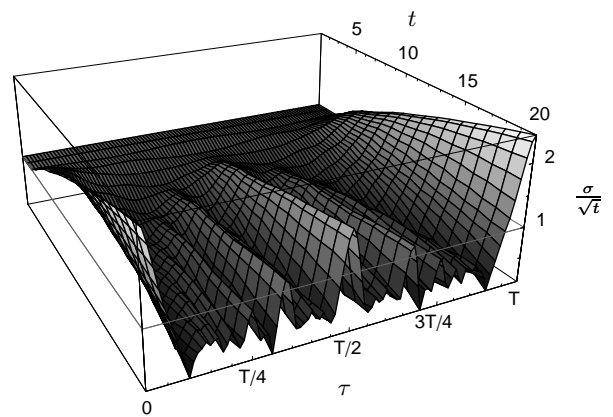}
 \caption{\label{var3Dnorm} Normalized standard deviation $\sigma / \sqrt{t}$ of stroboscopic Hadamard walks (harmonic oscillator model) for $t = 20$ steps with a time resolution of $\Delta \tau = \frac{T}{100}$. Sub- and superclassical spreading can clearly be seen.}
\end{figure}

We first observe that there is linear spreading of the walker's probability distribution on the line not just for the Hadamard walk case but also for certain time intervals $\tau$ after the initial oscillations have settled down (Fig.~\ref{meanvar}). Most interestingly some of these slower walks still outperform classical walks up to the simulated number of steps. However, for other values of $\tau$ the particle is tightly localised around the origin of the walk (Fig.~\ref{prob}). There are other intriguing features to be seen in the time evolution. Firstly the time intervals $\tau$ for which significant spreading occurs exhibit an interesting pattern (Fig.~\ref{varscan}) and appear to be symmetric about $\tau= \frac{T}{2}$. Comparing the whole range of stroboscopic walks (as $\tau$ is varied) even allows to distinguish between super- and subclassical spreading behaviour (Fig.~\ref{var3Dnorm}).

\begin{figure}
 \psfrag{X}[Bc][Bc]{$\tau$}
 \psfrag{Y}[Bc][Bc]{$\sigma$}
 \includegraphics{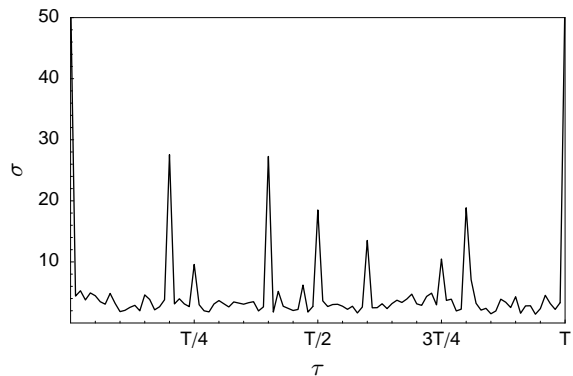}
 \caption{\label{fnrpvarscan} Standard deviation $\sigma$ of stroboscopic Hadamard walks (free non-relativistic particle model) after $t = 100$ steps with a time resolution of $\Delta \tau = \frac{T}{100}$.}
\end{figure}

In our discussion of numerical simulations we have merely focused on the harmonic oscillator Hamiltonian so far to point out the striking features of stroboscopic quantum walks. However, it is important to emphasize that this choice is just a special case of the Hamiltonians $\hat{H}_P$ allowed in Eq.~\ref{U} and can therefore only illustrate a fraction of the possible phenomena. In order to demonstrate this we also carried out numerical studies for the Hamiltonian $\hat{H}_P = \hat{p}^2/2m$ describing a free non-relativistic particle. Receiving discrete momentum kicks $\Delta p$ the particle walks in the space  $\mathcal{H}_P$ spanned by the discrete momentum eigenstates $\{ \ket{n \Delta p} \, \vert \, n \in \mathbb{Z} \}$, the Talbot time is given by $T = \frac{\pi m \hbar}{\Delta p^2}$. The results for this family of stroboscopic walks can be seen in Fig.~\ref{fnrpvarscan}.

Our results indicate that the concept of quantum walks may actually be embedded in a broader framework. The dynamics of the stroboscopic walks shows all the features of a nonlinear dynamical system in the presence of periodic perturbation. This extends from dynamical localisation -- spreading more slowly than in the classical case -- to linear diffusion as the period of the perturbation is varied. In particular, dynamical localisation of the particle seems to emerge if $\tau$ is not a rational multiple of the Talbot time. If on the other hand $\tau = p/q \, T$, where $p$ and $q$ are integers, we observe numerically a linear growth of the standard deviation in the asymptotic limit of a large number of steps, in other words the system exhibits a quantum resonance of order $q$. This behaviour is in direct analogy to the $\delta$-kicked rotator and has been extensively studied for this model \cite{Rei92, Izr79}. There is also a strong link from stroboscopic walks to recent experiments dealing with the slightly different $\delta$-kicked accelerator which exhibits quantum accelerator modes at Talbot time and certain fractions of it \cite{Obe99, Arc01, Sch03}. This connection to the field of quantum chaos is also supported by the recent discovery that the asymptotic behaviour of generalized walks is related to the $\delta$-kicked rotator \cite{Rom04}. As was first pointed out in \cite{Fis82} dynamical localisation in periodically kicked quantum systems is related to Anderson localisation of electronic states in disordered lattices \cite{And58, Kra93}. In the case of stroboscopic quantum walks the disorder enters via pseudorandom phase differences between components of the total quantum state except at Talbot time or rational multiples of it. In the context of Anderson localisation it has been argued though that pseudorandomness of the lattice is already sufficient to create localisation \cite{Gre84, Gri88}. This is precisely what we see in our simulations of stroboscopic quantum walks. If the time intervals between successive steps in our model are truly randomized (even slightly) we expect linear spreading to vanish completely.

In summary we have recast the coined quantum walks into a picture of periodic perturbations to an otherwise freely evolving system and introduced the model of a stroboscopic quantum walk. Discovering high order quantum resonances in our model we have also established the close link to nonlinear $\delta$-kicked systems in quantum chaos theory.

\begin{acknowledgments}
We thank Zhao-Yuan Ma, Salvador Venegas, Jacob Dunningham, Robert Schumann and Mark Lee for their help and insight. This work was supported by the UK Engineering and Physical Sciences Research Council, The Royal Society and Wolfson Foundation. O.~B. also thanks St John's College and Stiftung Maximilianeum for a student exchange place.
\end{acknowledgments}

\bibliography{Paper}

\end{document}